\documentclass[twoside,english]{iopart}
\usepackage[T1]{fontenc}
\usepackage[latin9]{inputenc}
\usepackage{geometry}
\usepackage{color}
\usepackage{array}
\usepackage{booktabs}
\usepackage{graphicx}

\makeatletter

\providecommand{\tabularnewline}{\\}

\usepackage{iopams}
\usepackage{setstack}

\usepackage[numbers, sort&compress]{natbib}

\newcommand{\eqref}[1]{(\ref{#1})}

\usepackage[singlelinecheck=on,justification=justified]{caption}
\usepackage[load-configurations=abbreviations,range-phrase = ~-~,range-units = single,separate-uncertainty = true]{siunitx}
\usepackage[siunitx]{circuitikz}
\RequirePackage{subscript}
\usepackage{ifpdf}
\ifpdf
\usepackage{epstopdf}
\fi
\usepackage[labelfont=bf]{caption}
\newcommand{\sym}[2]{\ensuremath{#1_\textrm{#2}}}

\newcommand{\n}[1]{\textrm{#1}}

\makeatother

\usepackage{babel}
\begin{document}

\title[Electro-Mechanical Properties of various Industrial Coated Conductors]{Electro-Mechanical Properties of REBCO Coated Conductors from various
Industrial Manufacturers at 77~K, self-field and 4.2~K, 19~T.}

\author{{\Large{}C Barth, G Mondonico and C Senatore}}

\address{{\large{}Department of Condensed Matter Physics (DPMC), University
of Geneva, Switzerland}}

\ead{{\large{}christian.barth@unige.ch}}
\begin{abstract}
Rare-earth-barium-copper-oxide (REBCO) tapes are now available from
several industrial manufacturers and are very promising conductors
in high field applications. Due to diverging materials and deposition
processes, these manufacturers' tapes can be expected to differ in
their electro-mechanical and mechanical properties. For magnets designers,
these are together with the conductors' in-field critical current
performance of the highest importance in choosing a suitable conductor.
In this work, the strain and stress dependence of the current carrying
capabilities as well as the stress and strain correlation are investigated
for commercial coated conductors from Bruker HTS, Fujikura, SuNAM,
SuperOx and SuperPower at \SI{77}{\kelvin}, self-field and \SI{4.2}{\kelvin},
\SI{19}{\tesla}.
\end{abstract}


\submitto{\SUST }

\maketitle

\section{Introduction}

Due to their high current density and zero direct current resistivity,
superconductors are the enabling component of powerful and efficient
magnets for accelerators, nuclear magnetic resonance (NMR), magnetic
resonance imaging (MRI) and fusion reactors. As the superconductors'
current carrying capabilities are not only influenced by the magnetic
field and the operating temperature but also by the mechanical strain
\cite{Kasaba2001,ZHANG2008,Bruzzone2011}, management of the electromagnetic
forces is critical. Especially rare-earth-barium-copper-oxide (REBCO)
coated conductors are further pushing the boundaries of superconductivity
towards devices with unparalleled magnetic fields and operating current
densities. Ongoing high-field projects based on REBCO coils include
high-resolution NMR spectrometers at \SI{1.3}{\giga\hertz}, corresponding
to a magnetic field of \SI{30.5}{\tesla} \cite{Bascunan2014,Senatore2014},
dipole magnets at \SI{20}{\tesla} in the perspective of particle
colliders with 10 times the present LHC energy \cite{Eucard2-temp},
the DEMO design studies for the future thermonuclear fusion power
plants \cite{Gade2014}. At these applications' extreme boundary conditions,
the conductors have to withstand enormous longitudinal strains and
stresses (\SI{\approx220}{\mega\pascal} in Bi2223 NMR coils \cite{Kiyoshi2011},
\SI{\approx330}{\mega\pascal} in  HTS fusion magnets \cite{Bansal2008} and \SI{\approx500}{\mega\pascal}
in REBCO NMR coils \cite{Otsuka2010}) making the electro-mechanical
and mechanical properties as well as the limits of irreversibility
of the highest importance for the magnets' design. In transverse direction
(parallel to c-axis), REBCO tapes are stress sensitive with irreversibility
limits below \SI{30}{\mega\pascal} \cite{Laan2007,Senatore2014}.
This low c-axis strength can be limiting factor in REBCO rotating
machinery or wet wound coils \cite{Takematsu2010,Barth2013-epoxy}.

Studies of the electrical transport properties under longitudinal
mechanical loads are routinely performed on Nb\textsubscript{3}Sn
wires \cite{Kock1977,Ekin1995,Seeber2007,Jewell2010,Goodrich2011}
and cables \cite{yukikazu2009,Liu2011,Liu2011a}, where strong effects
of applied strain on the critical current are observed even before
that irreversible degradation occurs. The behavior of REBCO tapes
under strain has been mainly investigated at 77 K in self-field \cite{Cheggour2005,Shin2005,Uglietti2006,Sugano2010,VanderLaan2010,Shin2012},
while there are only very few electromechanical data at low temperature
/ high field \cite{Uglietti2006,Sunwong2013}. Moreover, industrial
manufacturers adopt combinations of different materials of substrate,
buffers, stabilizer and laminated support and this may lead to large
variations in the response of the tape to the applied strain.

In this work, we will compare the influence of mechanical strains
and stresses on the current carrying capabilities of REBCO coated
conductor tapes from five major industrial manufacturers. Experiments
have been carried out using the Walters spring (WASP) probe developed
at the University of Geneva \cite{Uglietti2003}. \SI{77}{\kelvin},
self-field and \SI{4.2}{\kelvin}, \SI{19}{\tesla} are chosen as
the measurements\textquoteright{} boundary conditions for the results
to be relevant for the low field, high temperature applications such
as cables, rotating machines or transformers as well as the high fields,
low temperature applications such as NMR magnets, inserts of high
field magnets or accelerator magnets. As magnets\textquoteright{}
electromagnetic Lorentz forces are mainly translated into hoop stresses
affecting the conductors longitudinally in tension, this direction
of mechanical loads is used in the following investigations. The paper
first discusses the peculiarities of WASP measurements on REBCO tapes
(section~\ref{sec:REBCO-Walter-spring}). Section~\ref{sec:Experimental-details}
is devoted to the description of the measurement procedures and of
the investigated samples. The results of the experiments are reported
in section~\ref{sec:Results} and discussed in section~\ref{sec:Discussion-and-conclusion}.

\section{Walter spring measurements of REBCO tapes\label{sec:REBCO-Walter-spring}}

\begin{figure}[!tbph]
\begin{centering}
\includegraphics[width=0.6\textwidth]{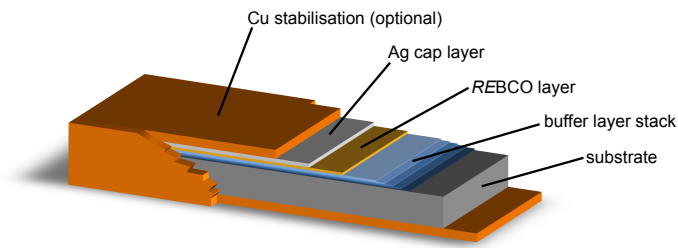}
\par\end{centering}

\caption{Schematic layout of REBCO tapes. The layers are not up to scale.\label{fig:REBCO-layout}}

\end{figure}

Rare-earth-barium-copper oxide tapes are layered conductors. By chemical
or physical means, layers of metal oxides (buffer layers), a superconducting
REBCO layer and stabilizing silver and copper layer are deposited
with a two dimensional texture upon a substrate of structural material
(usually Hastelloy\textsuperscript{\textregistered}, stainless steel
or nickel alloys). Due to this type of manufacturing, REBCO tapes
are often referred to as coated conductors. Their layout is shown
schematically in figure~\ref{fig:REBCO-layout}.

In this work, we measured the electro-mechanical properties of REBCO
tapes using a Walters spring \cite{Walters1986-Walter-Spring,seeber2012},
combining high sample lengths with the possibility to measure in tension
and in compression. This method however requires the samples to be
soldered to the spring, pre-straining them upon cool-down and thus
making a determination of their zero strain value necessary. Compared
to the other low temperature superconductors, there are significant
differences when performing Walters spring measurements on REBCO tapes.
The below mentioned solutions are verified and utilized within this
work.

\subsection{Mounting of the samples\label{sub:REBCO-mounting-procedure}}

The sample has to be soldered to the spring in order to transfer the
mechanical strain. With low temperature superconductors, this is commonly
done with \textcolor{black}{\textcolor{black}{Sn\textsubscript{50}Pb\textsubscript{32}Cd\textsubscript{18}}}
solder (melting point of \SI{145}{\celsius}) using a soldering iron
at \SIrange{350}{400}{\celsius} and heaters (\SIrange{100}{120}{\celsius}
set-point) attached to the spring. With this setup, the heat transferred
from the soldering iron through the sample is sufficient to locally
heat the spring above the solder's melting point in order to bond
the sample \cite{Uglietti2005}.
\begin{description}
\item [{Challenges~with~REBCO~tapes:}] REBCO coated conductors however
exhibit very low transverse thermal conductivity \cite[chap. 4.3.4]{Barth-PhD}
(roughly three orders of magnitude lower than in tape direction \cite{Bonura2014a,Bonura2014b})
requiring higher soldering iron temperatures and leading to temperatures
\ensuremath{\gg\SI{200}{\celsius}} on the coated conductor surface.
At these temperatures, the oxygen content in the superconducting layer
is reduced, resulting in a degradation of the current carrying capabilities
\cite{Kim2009}. As the soldering is done by hand, along the length
of the \SI{1}{\metre} long sample, the temperatures and the exposure
times vary. Therefore, the critical current degradation also varies
along the length, yielding very inhomogeneous samples and preventing
precise measurements. REBCO coated conductors thus require a different
mounting method.
\item [{Solution~for~REBCO~tapes:}] As it is necessary for the soldering
of the sample to heat the spring above the melting point of the solder,
significant heat input is required. In order to prevent the sample's
current carrying capabilities from becoming inhomogeneous during mounting,
the temperature and exposure times have to be constant along the length
of the coated conductor sample. This is achieved with a two step approach:
in the first step, the spring and the REBCO tape sample are coated
with\textcolor{red}{{} }\textcolor{black}{\textcolor{black}{Sn\textsubscript{50}Pb\textsubscript{32}Cd\textsubscript{18}}}
solder (melting point of \SI{145}{\celsius}, containing a colophony
base flux core). The Walters spring is nickel coated and can be easily
coated with solder by using stainless steel flux. By attaching the
REBCO tape vertically to a surface with low thermal conductivity and
using a soldering iron at \SI{180}{\celsius}, the soldering is quick
and without any critical current degradation. The REBCO sample is
then wound around the spring, coated side inwards. Two pieces of adhesive
tape keep the sample in place. In the second step, the sample and
the spring are put into a vacuum furnace and heated to \SI{220}{\celsius}
for \SI{1.5}{\hour}. The heat treatment in the vacuum furnace reduces
the sample's critical current. The reduction however is homogeneous
along the length of the sample. The reduction depends on the length
of the heat treatment, normalized critical currents are therefore
used in all following investigations. An impact of the heat treatment
on the sample's electro-mechanical properties is investigated in subsection~\ref{sub:Impact-of-the-heat-treatment}.
\end{description}

\subsection{Determination of zero strain\label{sub:Zero-strain-determination}}

As the samples are soldered to the spring, they follow the spring's
thermal expansion upon cool-down. The thermal expansion mismatch between
the spring (in our case titanium-aluminum-vanadium Ti-6Al-4V) and
the sample, induces pre-strain which has to be determined and subtracted.
With an LTS sample this is done in a two step procedure. In a first
step the sample is soldered at the spring's ends while the rest of
the sample remains ``free'' it follows its own thermal expansion
during cool-down and is without pre-strain \cite{Seeber2005}. The
sample's critical current is now determined at a magnetic field and
temperature at which the measurements is to take place. In a second
step, the soldering of the sample is completed, connecting it fully
to the spring. The sample's current carrying capabilities are measured
at the previously used temperature and field values, while slowly
turning the spring until the same critical current is obtained. This
position of the spring corresponds to the sample's pre-strain and
is usually referred to as its zero applied strain value. For low temperature
superconductors, especially for Nb\textsubscript{3}Sn, the variation
of critical current with strain, the ``strain effect'' is very pronounced,
making this method precise and reliable. Critical currents from different
fields can be used in order to further enhance the accuracy.
\begin{description}
\item [{Challenges~with~REBCO~tapes:}] For REBCO coated conductors on
the other hand, the strain effect is significantly lower compared
with low temperature superconductors. Within the reversible region,
mechanical strains reduce the current carrying capabilities solely
by a few percent, preventing precise determination of the zero strain
with the above mentioned procedure. Electro-mechanical Walters spring
measurements with REBCO tapes require therefore a different mean to
obtain their point of zero strain.
\item [{Solution~for~REBCO~tapes:}] The use of two strain gauges, one
glued onto the sample and one glued to a free piece of REBCO tape
located close to the spring, enables direct measurement of the strain
induced into the sample during cool-down \cite{Uglietti2003,Uglietti2006}.
\SI{350}{\ohm} strain gauges are used that are fed with a constant
current of \SI{1}{\milli\ampere} in series with a high precision
current source and are read in parallel with two Nano-volt meters.
The readout of the Nano-volt meters is zeroed at room temperature,
the probe is inserted into the cryostat and cooled down to measurement
temperature (\SI{77}{\kelvin}, \SI{4.2}{\kelvin} respectively).
The readouts diverge as the strain gauge onto the free piece of REBCO
tape follows the coated conductor's thermal expansion while the strain
gauge on the sample is mainly influenced by the thermal expansion
of the spring material. With \SI{-0.17}{\percent} at \SI{4.2}{\kelvin}
and \SI{-0.16}{\percent} at \SI{77}{\kelvin}, the thermal expansion
of the spring (Ti-6Al-4V) is very low, much lower than the coated
conductors' thermal expansion \cite{Barth2013-epoxy}. The REBCO tapes
samples are therefore strained in tensile direction. Slowly turning
the spring into compression reduces the differences of the strain
gauges' readouts until they are equalized around \SI{-0.14}{\percent}
to \SI{-0.15}{\percent} strain depending on the manufacturer of the
coated conductor sample. This is the zero strain value \sym{\epsilon}{zero}
of the sample at the used measurement temperature. With repeated measurements
of the same type of coated conductor sample, this method has proven
to be precise, reliable and repeatable. Furthermore, using the temperature
dependent gauge factor of the strain gauges, it is possible to use
the sample's gauge to directly calibrate the angle and applied strain
correlation of the Walters spring. This method to determine the zero
strain value is used in all following experiments.
\end{description}

\section{Experimental details\label{sec:Experimental-details}}

\begin{figure}[!tbph]
\begin{centering}
\includegraphics[width=0.5\textwidth]{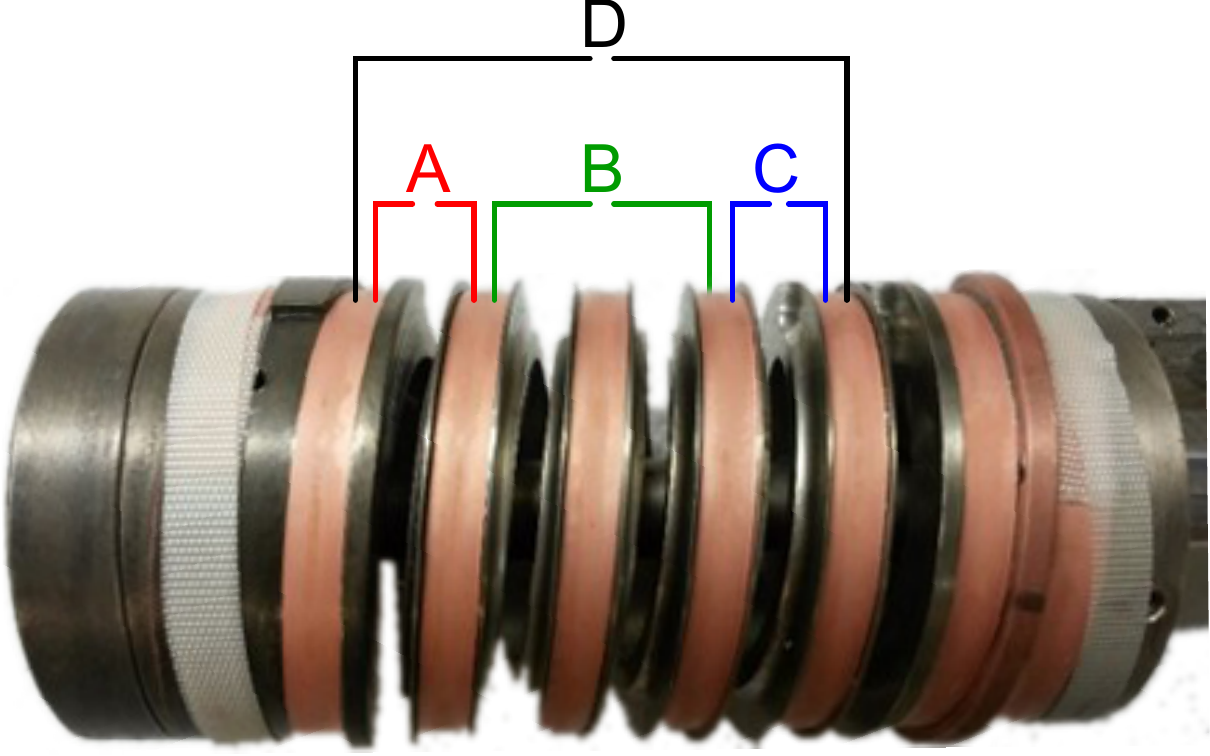}
\par\end{centering}

\caption{Walters spring and REBCO tape sample including the position of the
voltage taps.\label{fig:WASP+sample}}
\end{figure}

A four-turn, titanium-vanadium-aluminum (Ti-6Al-4V) Walters spring
with an active diameter of \SI{39}{\milli\metre} is used. There is
a \SI{4.1}{\milli\metre} wide grove for the sample. The total length
of the sample is \SI{0.985}{\metre}, there are \SI{0.486}{\metre}
of the sample on the spring usable for measurements. The Walters spring
and the mounted sample are shown in figure~\ref{fig:WASP+sample}.

\subsection{Determination of the critical current}

\begin{figure}[!tbph]
\begin{centering}
\includegraphics[width=0.5\textwidth]{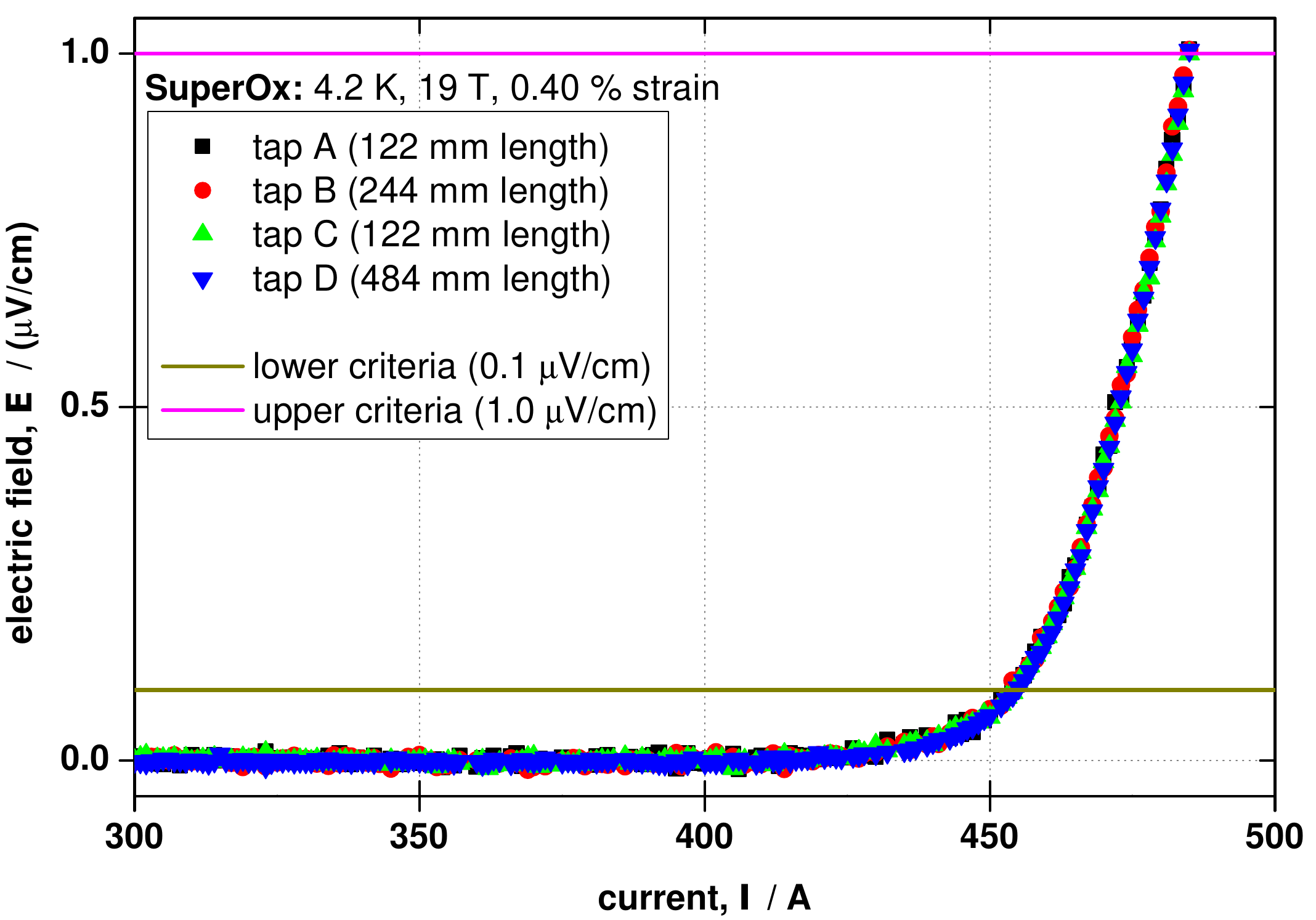}
\par\end{centering}

\caption{Exemplary electric field vs. current curve of the Walters spring measurement
system. The measurement is of SuperOx tape at \SI{4.2}{\kelvin},
\SI{19}{\tesla} at a \SI{0.40}{\percent} tensile strain.\label{fig:E-vs-I-curve}}
\end{figure}

Four pair of voltage taps are used to determine the samples' critical
current. One pair of voltage taps over the first turn (A: \SI{122}{\milli\metre}
measurement length) , one pair over the second and third turns (B:
\SI{244}{\milli\metre} measurement length) and one pair over the
fourth turn (C: \SI{122}{\milli\metre} measurement length). The final
pair of voltage taps is over the whole spring (D: \SI{484}{\milli\metre}
measurement length). During the measurement, the current is increased
in small steps while four Nano-volt meters simultaneously read the
voltage over the different turns of the sample. This arrangement allows
low noise measurements for precise determination of the critical currents
and the critical current index values (n-values). An exemplary electric
field vs. current curve, demonstrating the precision of the system,
is shown in figure~\ref{fig:E-vs-I-curve}. In all following experiments,
the critical current is defined as the current at the critical electric
field of \SI{0.1}{\micro\volt\per\centi\metre} (lower criteria) of
the voltage tap pair spanning the whole spring (D: \SI{484}{\milli\metre}
of measurement length).

\subsection{Determination of the strain dependence of the critical current\label{sub:Strain-determination}}

If the superconducting layer is not located in the sample's center,
strain is induced while bending around the spring. The bending strain
\sym{\epsilon}{bending} depends on the displacement of the superconducting
layer from the tape's neutral axis and is determined by comparing
the circumference of the neutral axis and the circumference of the
REBCO layer. In all investigated samples, the superconducting layer
is located closer to the center of the spring compared with the sample's
neutral axis, the superconducting layer is therefore compressed upon
bending around the spring. The sample's strain $\epsilon$ is obtained
by subtracting the zero strain value \sym{\epsilon}{zero} and the
bending strain \sym{\epsilon}{bending} from the applied strain. In
order to determine the samples' irreversibility limits (irreversible
strain \sym{\epsilon}{irr} and irreversible stress \sym{\sigma}{irr}),
after each critical current measurement $I_{\n{c}}(\epsilon_{\n{n}})$
at certain strain \sym{\epsilon}{n}, the strain is reduced to the
previous strain step \sym{\epsilon}{n-1} and the critical current
is measured again $I_{\n{c}}^{\n{back}}(\epsilon_{\n{n-1}})$. Within
the reversible strain region of the sample, the critical current recovers
yielding the same value as on the first measurement at this strain
$I_{\n{c}}^{\n{back}}(\epsilon_{\n{n-1}})=I_{\n{c}}(\epsilon_{\n{n-1}})$.
However, if the sample is damaged, this procedure is not reversible
and the critical current of the ``backstep'' is lower than the value
found at the first measurement $I_{\n{c}}^{\n{back}}(\epsilon_{\n{n-1}})<I_{\n{c}}(\epsilon_{\n{n-1}})$.
With this method, the irreversible strain \sym{\epsilon}{irr} of
the sample is therefore defined as the strain interval between the
highest strain step after which at least one of the backsteps recovered
and first strain step with no recovering backsteps. This interval
is converted into a stress interval using the stress and strain correlation
of the corresponding sample (subsection~\ref{sub:Stress-strain})
yielding the irreversible stress limit \sym{\sigma}{irr}. All following
experiments utilize this method to determine the samples' irreversibility
limits.

\subsection{Determination of the stress dependence of the critical current\label{sub:Stress-determination}}

In order to translate the strain behavior of the coated conductors
to stress, their stress - strain correlation is measured at the same
temperatures of the electro-mechanical tests using a free standing
mechanical measurement system. One side of the sample is fixed while
the other is attached to a pulling rod. Nyilas type extensometers
\cite{Nyilas2005a,Nyilas2005b} are attached to the sample reading
the strain while a high precision load cell located in the pulling
rod directly above the sample reads the stress. After cool-down to
operating temperature, the sample is slowly strained, releasing the
force to two thirds of its current value at \SI{0.3}{\percent} and
\SI{0.5}{\percent} strain. The slope of these so called ``mini loops''
give the sample's Young's modulus \sym{E}{0}. For the determination
of the stress, the cross sectional area of the sample is highly important,
thus the precise dimensions of the coated conductor sample are obtained
through micro-graphs. The stress versus strain curves are fitted with
high order polynomial functions in order to translate the measured
strains of the Walters spring experiments into mechanical stresses.

\subsection{Investigated samples}

In this work, REBCO coated conductors from the major industrial manufacturers
Bruker HST, Fujikura, SuNAM, SuperOx and SuperPower are characterized.
An overview of the investigated samples including their strain corrections,
zero strains \sym{\epsilon}{zero} and bending strains \sym{\epsilon}{bending}
is given in table~\ref{tab:Investigated-samples}.

\begin{table*}[!tbph]
\caption{Overview of the investigated REBCO tapes and their strain corrections:
bending strain \sym{\epsilon}{bending} and zero strain \sym{\epsilon}{zero}.
Tape width and thickness are obtained from micro-graphs. Engineering
current densities are from the samples as received (without heat treatment).
Values marked with {*} are extrapolated using scaling relations.\label{tab:Investigated-samples}}

\centering{}{\tiny{}}%
\begin{tabular*}{1\textwidth}{@{\extracolsep{\fill}}c>{\centering}p{0.1\textwidth}>{\centering}p{0.1\textwidth}>{\centering}p{0.1\textwidth}>{\centering}p{0.1\textwidth}>{\centering}p{0.1\textwidth}}
\toprule 
 & {\tiny{}Bruker HST} & {\tiny{}Fujikura} & {\tiny{}SuNAM} & {\tiny{}SuperOx} & {\tiny{}SuperPower}\tabularnewline
\midrule
\midrule 
{\tiny{}width} & {\tiny{}\SI{4.10}{\milli\metre}} & {\tiny{}\SI{3.05}{\milli\metre}} & {\tiny{}\SI{4.00}{\milli\metre}} & {\tiny{}\SI{4.04}{\milli\metre}} & {\tiny{}\SI{4.00}{\milli\metre}}\tabularnewline
\midrule
\midrule 
{\tiny{}thickness} & {\tiny{}\SI{153}{\micro\metre}} & {\tiny{}\SI{161}{\micro\metre}} & {\tiny{}\SI{110}{\micro\metre}} & {\tiny{}\SI{112}{\micro\metre}} & {\tiny{}\SI{101}{\micro\metre}}\tabularnewline
\midrule
\midrule 
{\tiny{}substrate} & {\tiny{}\SI{100}{\micro\metre} stainless steel} & {\tiny{}\SI{75}{\micro\metre} Hastelloy} & {\tiny{}\SI{60}{\micro\metre} Hastelloy} & {\tiny{}\SI{60}{\micro\metre} Hastelloy} & {\tiny{}\SI{50}{\micro\metre} Hastelloy}\tabularnewline
\midrule
\midrule 
{\tiny{}Cu stabilization} & {\tiny{}\ensuremath{2\times}\SI{15}{\micro\metre} electroplated} & {\tiny{}\ensuremath{1\times}\SI{75}{\micro\metre} laminated} & {\tiny{}\ensuremath{2\times}\SI{18}{\micro\metre} electroplated} & {\tiny{}\ensuremath{2\times}\SI{20}{\micro\metre} electroplated} & {\tiny{}\ensuremath{2\times}\SI{20}{\micro\metre} electroplated}\tabularnewline
\midrule
\midrule 
{\tiny{}\ensuremath{J_{\n{eng}}(\SI{77}{\kelvin}, \n{self-field})}} & {\tiny{}\SI{87}{\ampere\per\milli\metre\squared}} & {\tiny{}\SI{350}{\ampere\per\milli\metre\squared}} & {\tiny{}\SI{445}{\ampere\per\milli\metre\squared}} & {\tiny{}\SI{305}{\ampere\per\milli\metre\squared}} & {\tiny{}\SI{225}{\ampere\per\milli\metre\squared}}\tabularnewline
\midrule
\midrule 
{\tiny{}\ensuremath{J_{\n{eng}}(\SI{4.2}{\kelvin}, \SI{19}{\tesla}\; ||)}} & {\tiny{}\SI{2.0}{\kilo\ampere\per\milli\metre\squared}\textsubscript{*}} & {\tiny{}\SI{3.5}{\kilo\ampere\per\milli\metre\squared}\textsubscript{*}} & {\tiny{}\SI{2.0}{\kilo\ampere\per\milli\metre\squared}} & {\tiny{}\SI{1.1}{\kilo\ampere\per\milli\metre\squared}} & {\tiny{}\SI{3.9}{\kilo\ampere\per\milli\metre\squared}\textsubscript{*}}\tabularnewline
\midrule
\midrule 
{\tiny{}\sym{\epsilon}{bending}} & {\tiny{}\SI{0.030}{\percent}} & {\tiny{}\SI{0.002}{\percent}} & {\tiny{}\SI{0.018}{\percent}} & {\tiny{}\SI{0.017}{\percent}} & {\tiny{}\SI{0.014}{\percent}}\tabularnewline
\midrule
\midrule 
{\tiny{}\ensuremath{\epsilon_{\n{zero}}(\SI{77}{\kelvin})}} & {\tiny{}\SI{0.145}{\percent}} & {\tiny{}\SI{0.139}{\percent}} & {\tiny{}\SI{0.142}{\percent}} & {\tiny{}\SI{0.138}{\percent}} & {\tiny{}\SI{0.142}{\percent}}\tabularnewline
\midrule
\midrule 
{\tiny{}\ensuremath{\epsilon_{\n{zero}}(\SI{4.2}{\kelvin})}} & {\tiny{}\SI{0.152}{\percent}} & {\tiny{}\SI{0.145}{\percent}} & {\tiny{}\SI{0.148}{\percent}} & {\tiny{}\SI{0.144}{\percent}} & {\tiny{}\SI{0.149}{\percent}}\tabularnewline
\bottomrule
\end{tabular*}
\end{table*}

\subsection{Impact of the heat treatment on the electro-mechanical properties\label{sub:Impact-of-the-heat-treatment}}

\begin{figure*}[!tbph]
\begin{centering}
\includegraphics[width=0.5\textwidth]{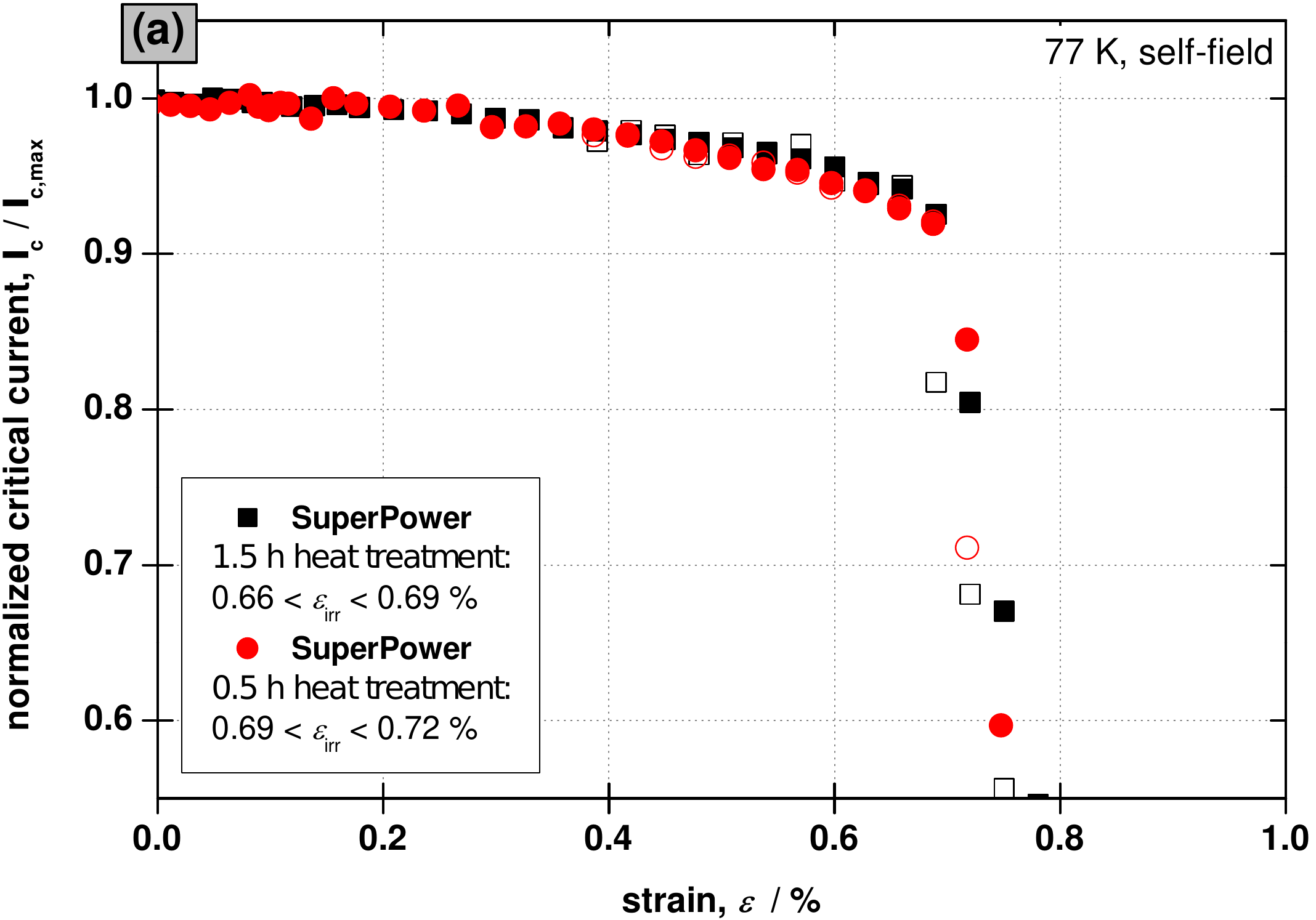}\includegraphics[width=0.5\textwidth]{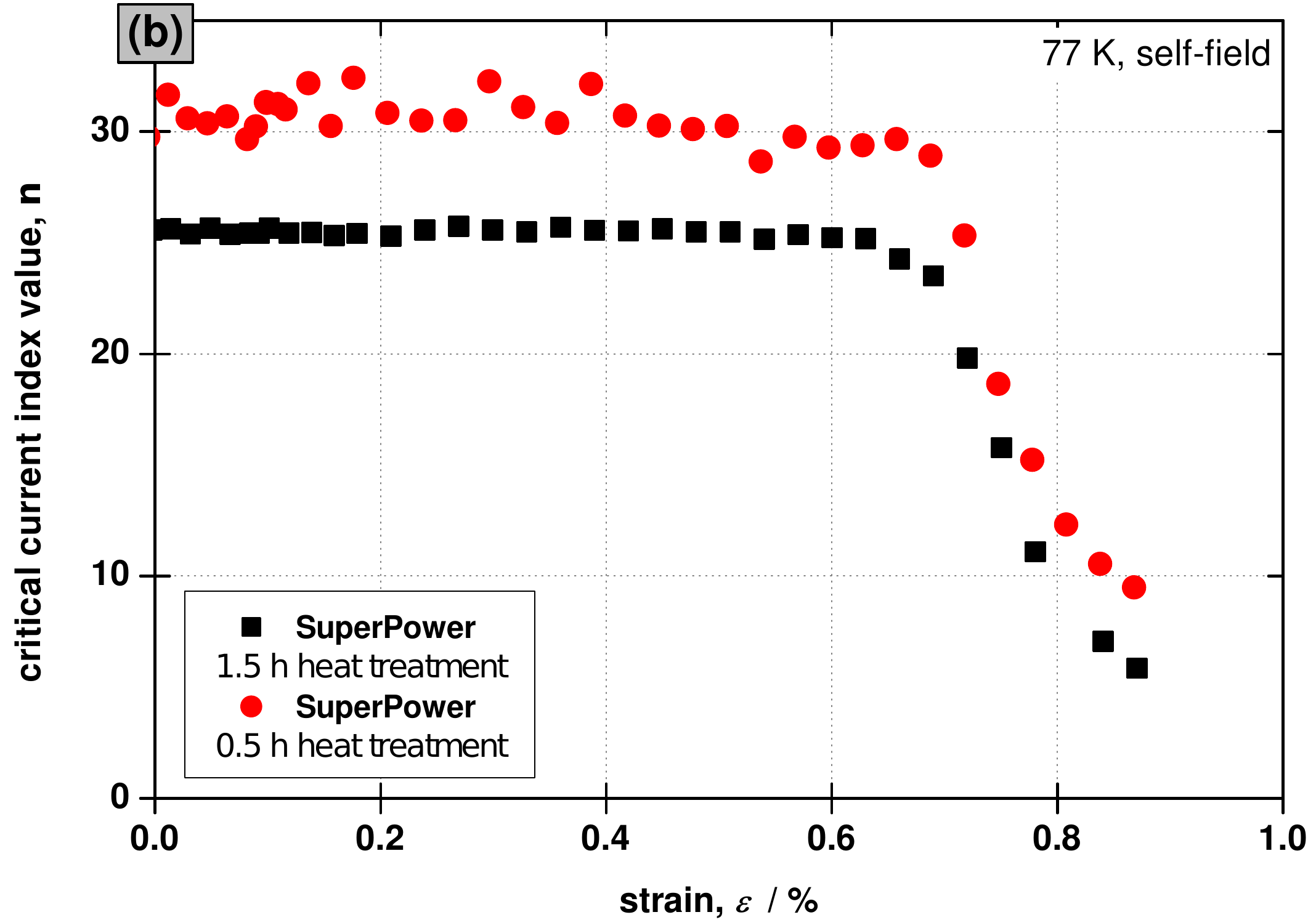}
\par\end{centering}

\caption{Characterization of the electro-mechanical properties of SuperPower
REBCO tapes with different heat treatment times (long heat treatment
of \SI{1.5}{\hour} in black and short heat treatment of \SI{0.5}{\hour}
in red). The strain dependence of the normalized critical current
is shown in the graph (a) and the strain dependence of the critical
current index in graph (b). Measurements are performed at \SI{77}{\kelvin},
self-field.\label{fig:Influence-heat-treatment}}
\end{figure*}

As the samples are heated in a vacuum furnace to \SI{220}{\celsius}
during mounting, their superconducting layers are partially depleted
of oxygen with the level of depletion depending on the duration of
the heat treatment. In order to investigate the influence of the oxygen
depletion on the electro-mechanical properties, two samples of the
same REBCO tapes are mounted using two different heat treatment times
(long heat treatment of \SI{1.5}{\hour} and short heat treatment
of \SI{0.5}{\hour}) . Their electro-mechanical properties are measured
in a liquid nitrogen bath at \SI{77}{\kelvin}, self-field. The strain
dependence of their normalized critical current as well as their irreversible
strain \sym{\epsilon}{irr} limits are shown in graph (a) of figure~\ref{fig:Influence-heat-treatment}
while graph (b) gives the strain dependence of their critical current
index values (n-values) $n$. Both samples behave identically, their
normalized critical current versus strain curves overlap and their
reversible strain limits \sym{\epsilon}{irr} are within \SIrange{0.66}{0.69}{\percent}
for the long heat treatment and within \SIrange{0.69}{0.72}{\percent}
for the short heat treatment. This is within one step of strain (\SI{0.03}{\percent})
as used for these measurements. The critical current index value (n-value)
versus strain curves are of the same shape, with higher n-values in
the case of the shorter heat treatment. This is to be expected as
the oxygen depletion of this sample is lower, resulting in higher
critical currents and therefore also higher n-values due to the strong
correlation between pinning properties, critical current and n-value.
Furthermore, these measurements demonstrate the reliability and repeatability
of the Walters spring setup, and show that any major impact of the
heat treatment on the electro-mechanical properties of the REBCO tape
samples can be excluded.

\section{Results\label{sec:Results}}

With the above mentioned setup and measurement procedure, all REBCO
tape samples (SuperPower, Bruker HTS, SuNAM, Fujikura and SuperOx)
are electro-mechanically characterized at \SI{77}{\kelvin}, self-field
and \SI{4.2}{\kelvin}, \SI{19}{\tesla}. The magnetic field is oriented
parallel to the tape surface. All samples have been subjected to the
same heat treatment during mounting (\SI{1.5}{\hour} at \SI{220}{\celsius}).
Their strain behavior (subsection~\ref{sub:Strain-dependence}) is
combined with stress-strain measurements at the same temperatures
(subsection~\ref{sub:Stress-strain}) in order to gain the samples\textquoteright{}
stress behavior (subsection~\ref{sub:sub:Stess-dependence}). To
ease the comparison of the results for the various manufacturers and
due to the critical current degradation during their heat treatments,
the current carrying capabilities of all coated conductor samples
are normalized to their maximal critical current.

\subsection{Normalized critical current vs. strain\label{sub:Strain-dependence}}

\begin{figure*}[!tbph]
\begin{centering}
\includegraphics[width=0.5\textwidth]{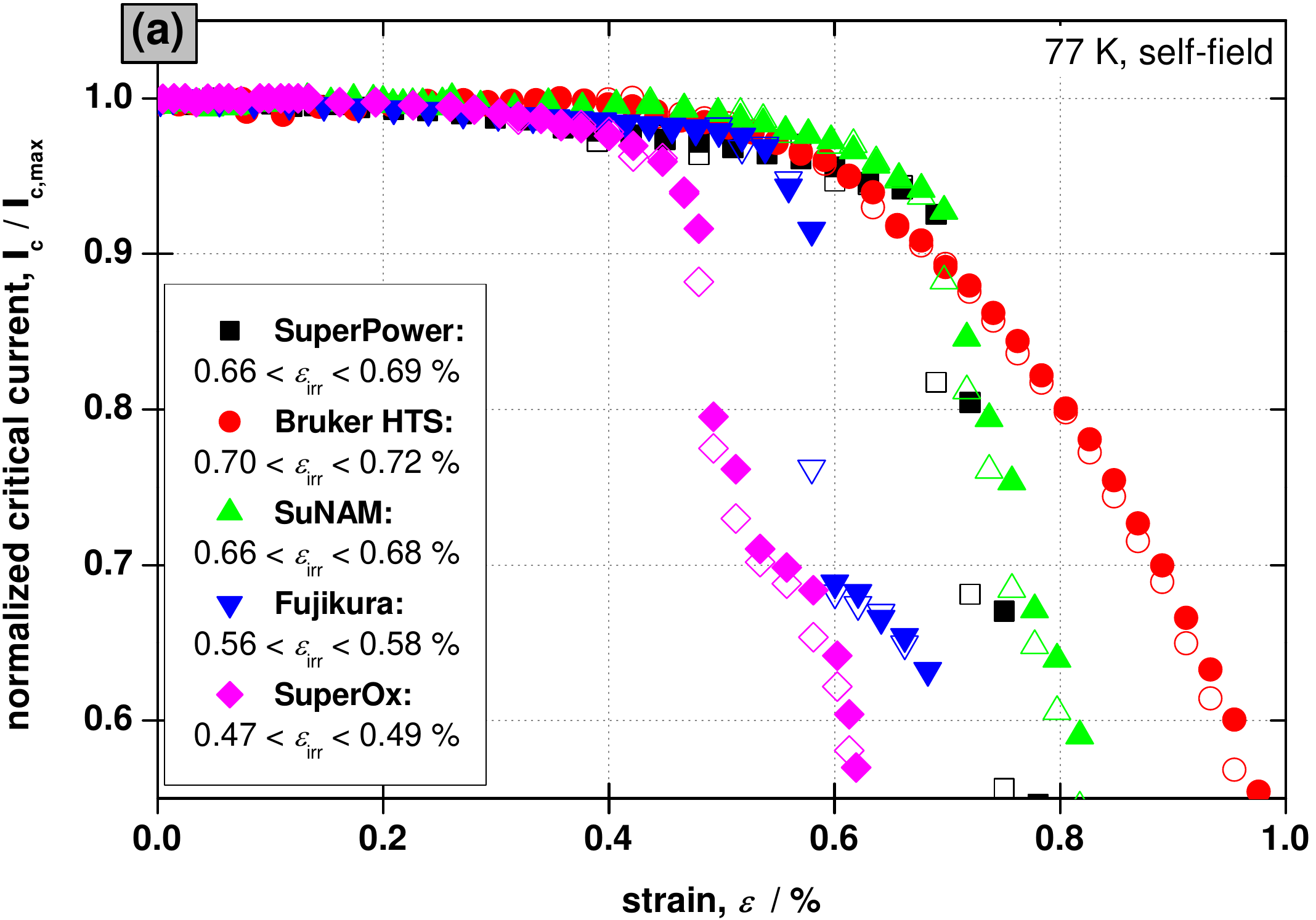}\includegraphics[width=0.5\textwidth]{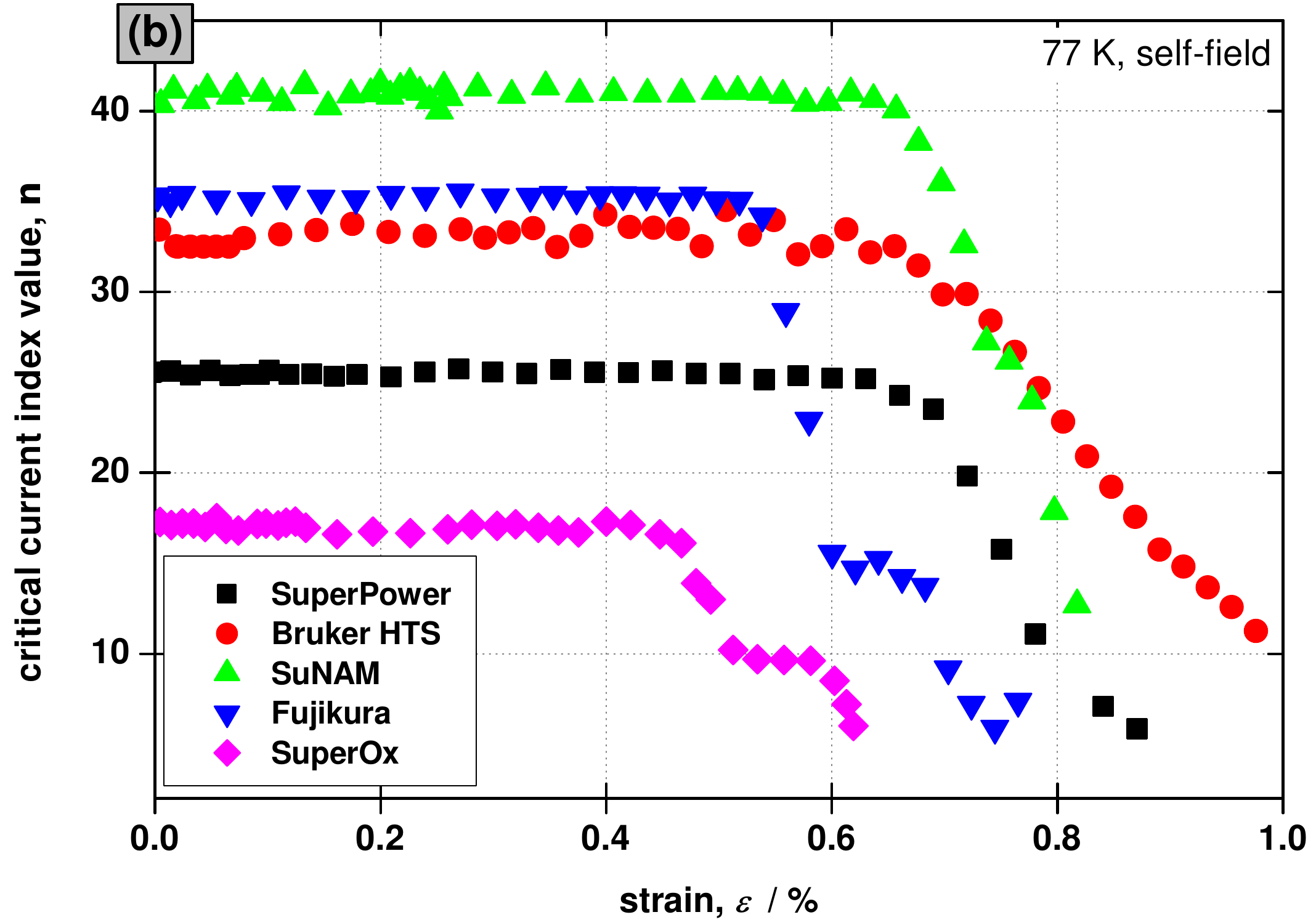}
\par\end{centering}

\caption{Strain dependence of various industrial REBCO tapes at \SI{77}{\kelvin},
self-field. Normalized critical current $I_{\n{c}}/I_{\n{c,max}}$
versus strain $\epsilon$ and irreversible strain limits \sym{\epsilon}{irr}
in graph (a). Critical current index value (n-value) versus strain
$\epsilon$ in the graph (b)\label{fig:Strain-dependence-77K}}
\end{figure*}

The \SI{77}{\kelvin}, self-field strain dependence of the samples
is shown in figure~\ref{fig:Strain-dependence-77K} with the normalized
critical current versus strain dependence in graph (a) and the n-value
dependence in graph (b).

\begin{figure}[!tbph]
\begin{centering}
\includegraphics[width=0.7\textwidth]{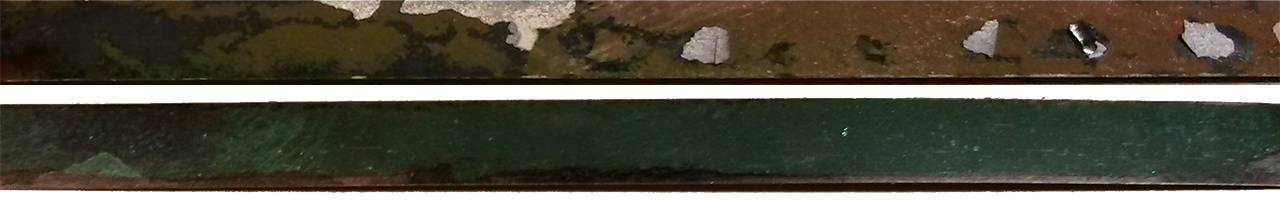}
\par\end{centering}

\caption{De-laminated Fujikura (top) and SuperOx (bottom) tapes after the electro-mechanical
Walters spring measurements. The coated conductor tapes are split
between the buffer layer stack and the superconducting REBCO layer
making their red buffer layers (Fujikura) and green buffer layers
(SuperOx) are fully visible.\label{fig:Delamination}}
\end{figure}

The tapes from SuperPower and SuNAM exhibit very similar strain dependence
with irreversible strain limits \sym{\epsilon}{irr} of \SIrange{0.66}{0.69}{\percent}
for SuperPower tapes and \SIrange{0.66}{0.68}{\percent} for SuNAM
tapes. Their normalized critical current curves overlap and their
n-value curves are of similar shape with higher n-values for the SuNAM
tapes. Bruker HTS tapes exhibit a ``more round'' normalized versus
critical current curve, indicating a stronger strain effect accompanied
by very high irreversible strain limits \sym{\epsilon}{irr} of \SIrange{0.70}{0.72}{\percent}.
The samples from Fujikura and SuperOx on the other hand have a ``step-like''
transition from their reversible to irreversible region. At a rather
low strain (\SIrange{0.55}{0.58}{\percent} for Fujikura tapes and
\SIrange{0.47}{0.49}{\percent} for SuperOx tapes), their critical
current is irreversibly reduced and stabilized at lower currents.
At higher strains, there is a second strong degradation of their current
carrying capabilities. This ``step-like'' behavior is also visible
in the n-value versus strain dependence in graph (b). At \SI{77}{\kelvin},
self-field the samples with the highest irreversible strain limits
are from Bruker HTS, followed by SuperPower and SuNAM. The irreversible
strain limits of Fujikura and SuperOx are significantly lower due
to the ``step-like'' transition. Furthermore, the samples from Fujikura
and SuperOx are fully de-laminated after these measurements as shown
in figure~\ref{fig:Delamination}. The tapes are split between the
buffer layer stack and the superconducting REBCO layer. This behavior
is not observed for any other type of coated conductor tapes. The
de-lamination and the step like behavior may indicated a interfacial
weakness of Fujikura and SuperOx coated conductor tapes. 

\begin{figure*}[!tbph]
\begin{centering}
\includegraphics[width=0.5\textwidth]{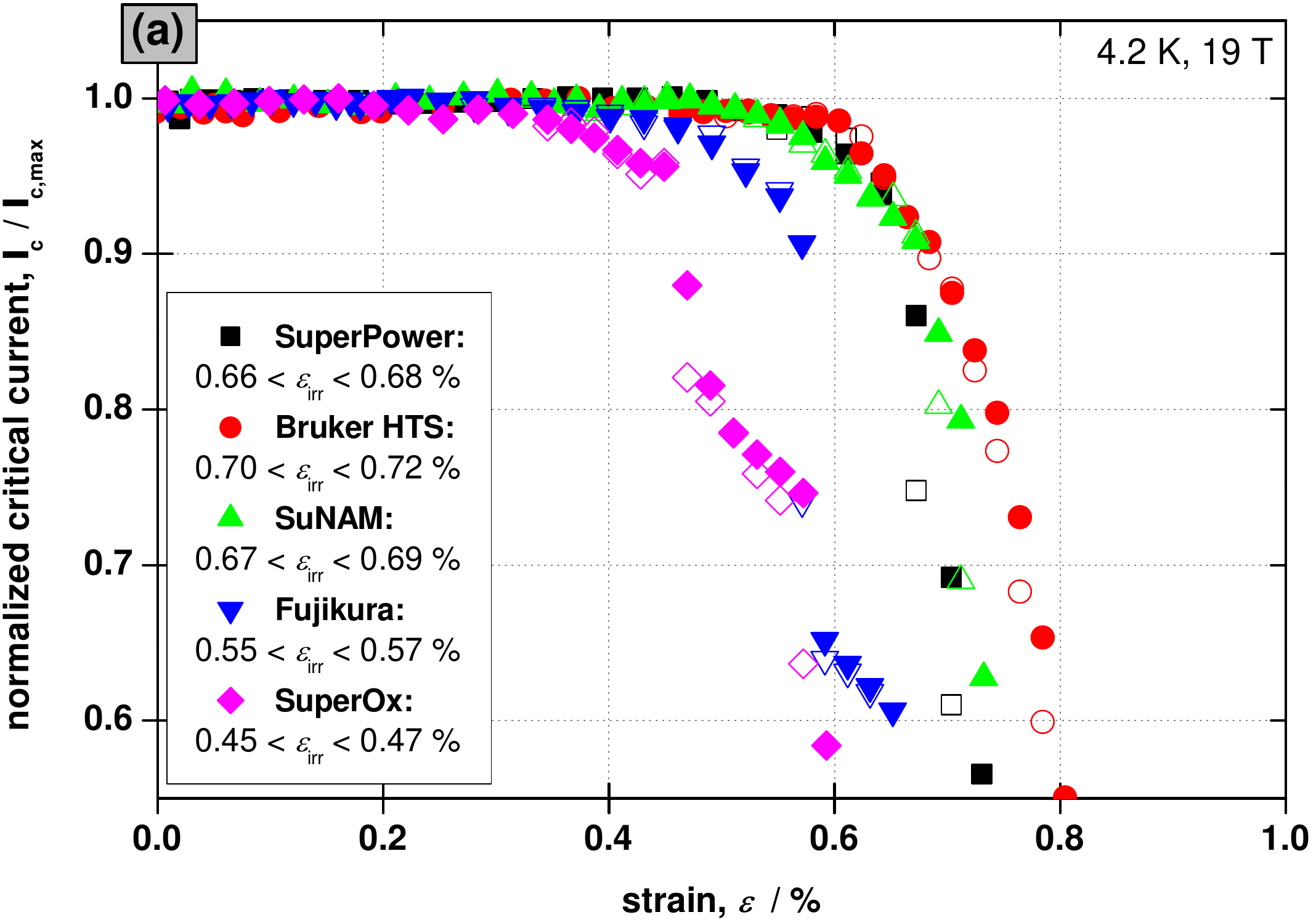}\includegraphics[width=0.5\textwidth]{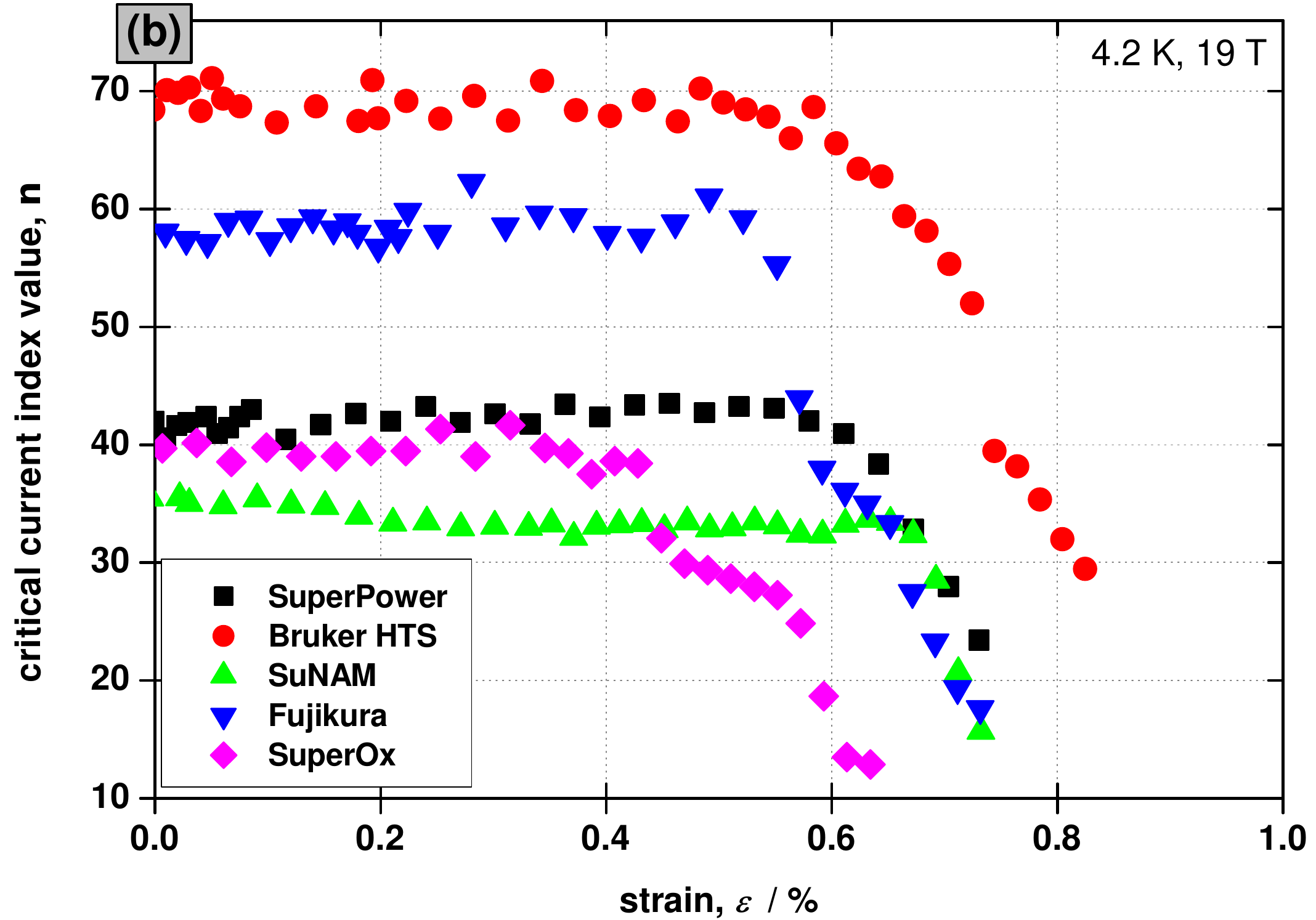}
\par\end{centering}

\caption{Strain dependence of various industrial REBCO tapes \SI{4.2}{\kelvin},
\SI{19}{\tesla}. Normalized critical current $I_{\n{c}}/I_{\n{c,max}}$
versus strain $\epsilon$ and irreversible strain limits \sym{\epsilon}{irr}
in the graph (a). Critical current index value (n-value) versus strain
$\epsilon$ in the graph in graph (b).\label{fig:Strain-dependence-4.2K}}
\end{figure*}

Similar behavior of the coated conductor samples is observed at \SI{4.2}{\kelvin},
\SI{19}{\tesla} as shown in figure~\ref{fig:Strain-dependence-4.2K};
normalized critical current versus strain in graph (a) and n-value
versus strain in graph (b). In the samples' reversible region, the
strain dependence of the normalized critical current is lower, the
transition to irreversibility however is steeper. The samples' irreversible
strain limits \sym{\epsilon}{irr} are similar; highest for Bruker
HTS tapes with \SIrange{0.70}{0.72}{\percent}, followed by SuNAM
with \SIrange{0.67}{0.69}{\percent} and SuperPower with \SIrange{0.66}{0.68}{\percent}.
Fujikura and SuperOx REBCO tapes exhibit significantly lower irreversible
strain limits with \SIrange{0.55}{0.57}{\percent} and \SIrange{0.45}{0.47}{\percent},
respectively. Full de-lamination of Fujikura and SuperOx tapes is
observed again. An impact of the mounting procedure cannot be excluded
as during the heat treatment the tapes are exposed to temperatures
\SI{10}{\percent} above the manufacturers' recommendations.

\subsection{Stress - strain correlation\label{sub:Stress-strain}}

\begin{figure*}[!tbph]
\begin{centering}
\includegraphics[width=0.5\textwidth]{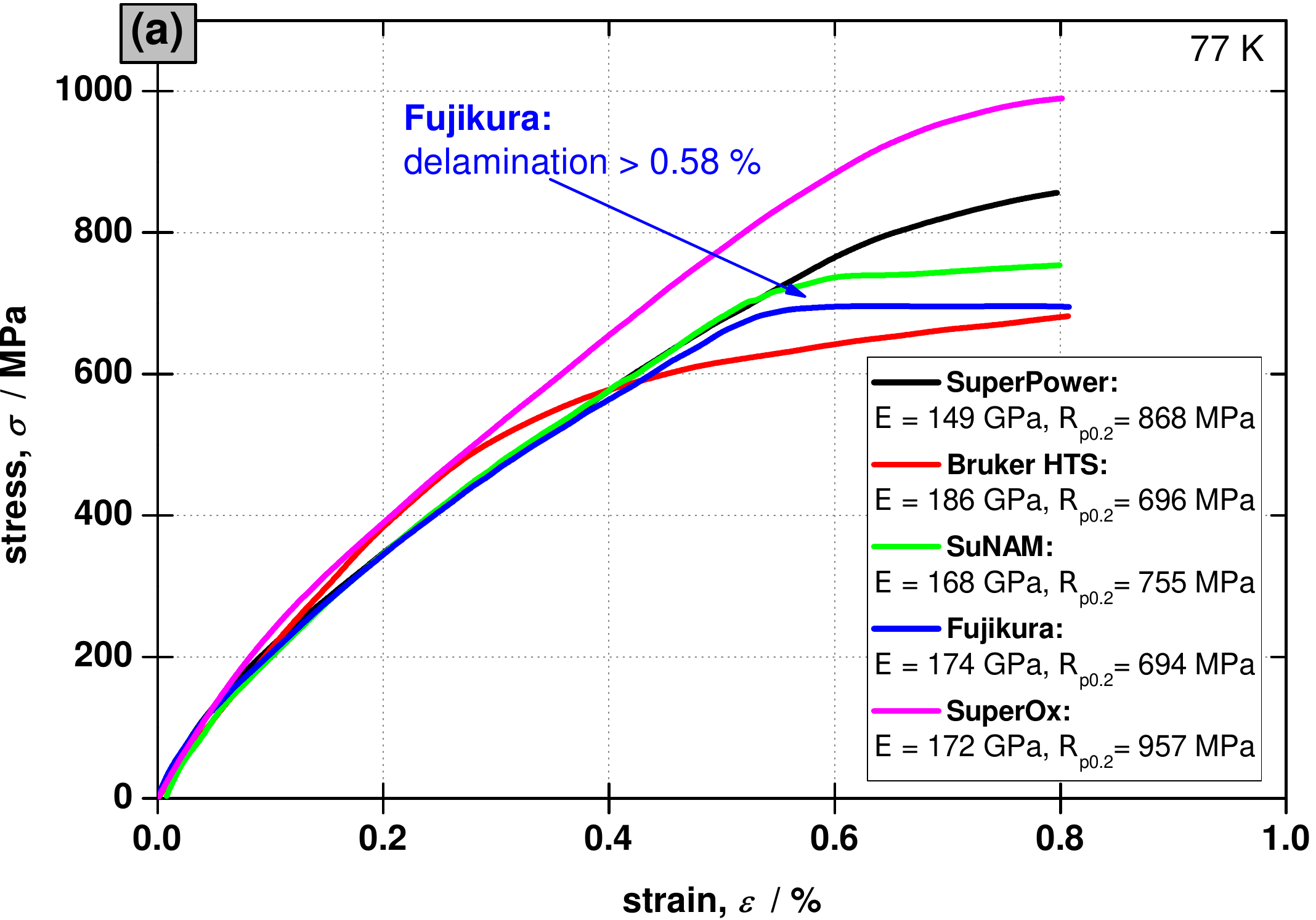}\includegraphics[width=0.5\textwidth]{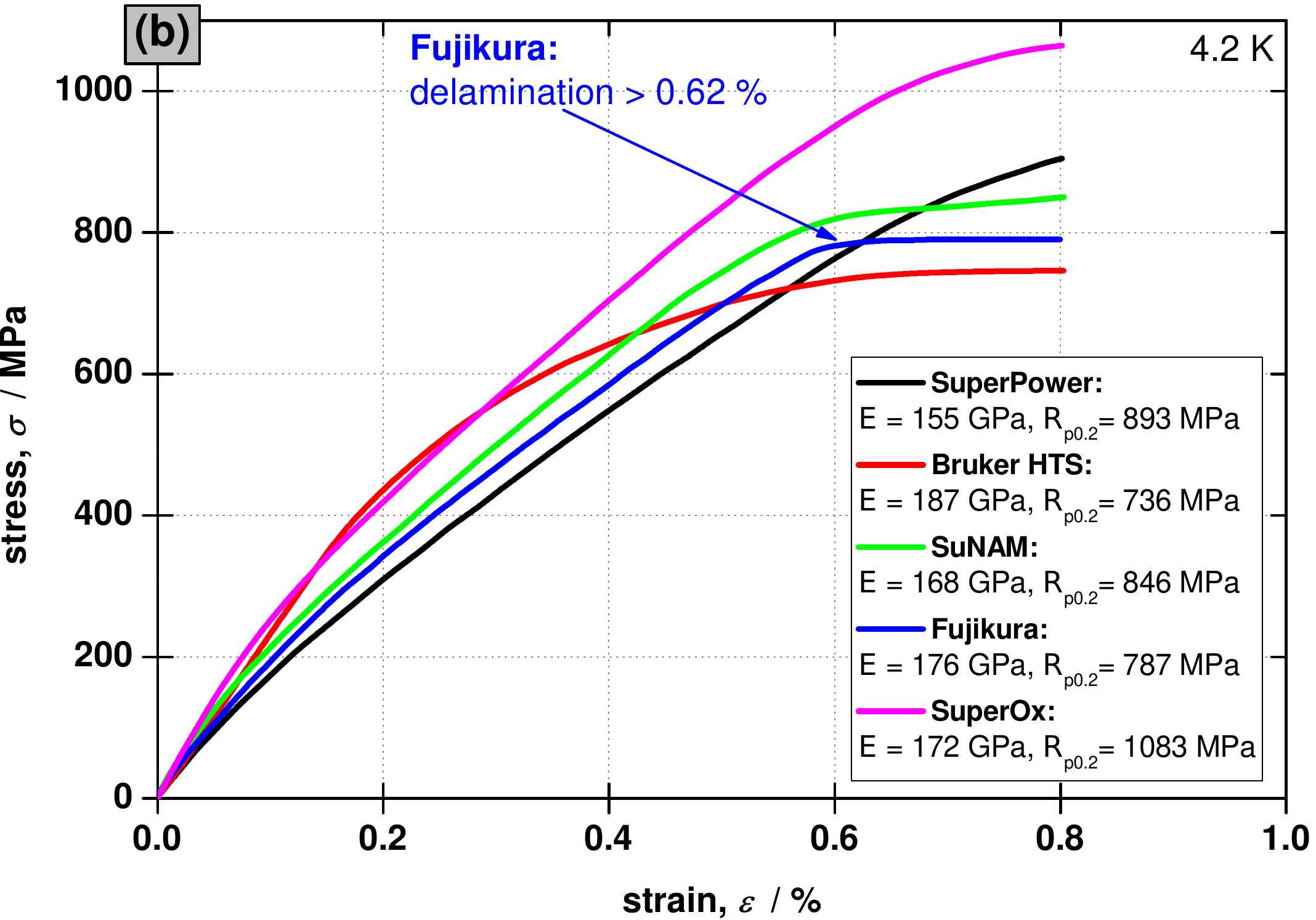}
\par\end{centering}

\caption{The stress - strain dependence of the various industrial REBCO tapes
is shown at \SI{77}{\kelvin} in graph (a) and at \SI{4.2}{\kelvin}
in graph (b). For the sake of readability, all ``mini loops'' (partial
release of the stress at certain strains) have been removed.\label{fig:Stress-strain}}
\end{figure*}

As described in subsection~\ref{sub:Stress-determination}, the samples'
correlation of applied strain and mechanical stress is measured at
\SI{77}{\kelvin} and \SI{4.2}{\kelvin}. The \SI{77}{\kelvin} data
is shown in graph (a) of figure~\ref{fig:Stress-strain} and the
\SI{4.2}{\kelvin} data in graph (b). For the sake of readability,
all ``mini loops'' (partial release of the stress at certain strains)
have been removed from the graphs. Young's moduli are obtained at
the ``mini loops'' performed at \SI{0.3}{\percent} strain. The
highest Young's modulus and yield strength are encountered for the
tape from SuperOx, followed by SuperPower and SuNAM. With its stainless
steel substrate, the stress - strain curve of the Bruker HTS tapes
is more ``round'' resulting in the poorest mechanical properties.
For the other manufacturers, the differences are mainly due to varying
copper - Hastelloy ratios. Comparing \SI{77}{\kelvin} and \SI{4.2}{\kelvin},
all sample's mechanical properties are very similar: their Young's
modulus are identical at both temperatures while their mechanical
yield strengths \sym{R}{p0.2} are roughly \SI{10}{\percent} higher
at \SI{4.2}{\kelvin}. All samples' stress-strain relations are slightly
nonlinear even at low strains. There are no distinct elastic and inelastic
regimes visible as a significant fraction of all samples, the copper,
is already in yielding solely due to the cool-down to cryogenic temperatures
\cite{EFDA-material-data-2007}.

\begin{figure}[!tbph]
\begin{centering}
\includegraphics[width=0.7\textwidth]{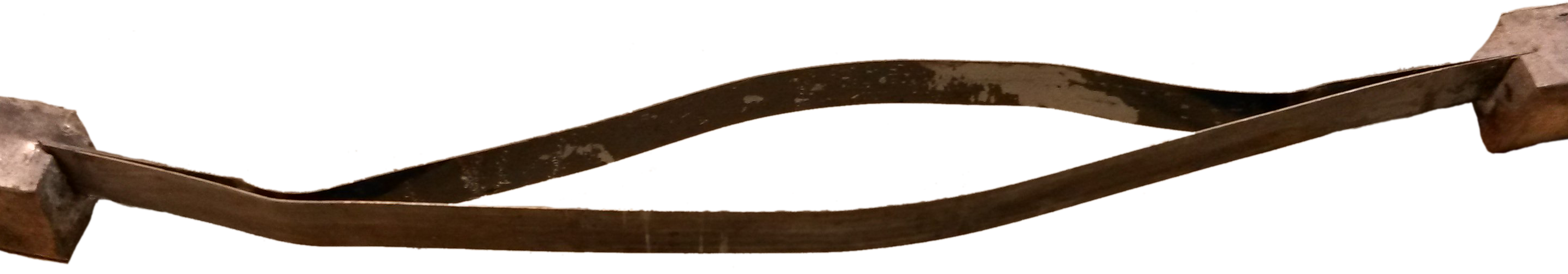}
\par\end{centering}

\caption{De-laminated Fujikura coated conductor samples after the stress -
strain measurements.\label{fig:Delamination-stress-strain} }

\end{figure}

The Fujikura samples fully de-laminate during the stress - strain
measurements as shown in figure~~\ref{fig:Delamination-stress-strain}.
At \SI{77}{\kelvin} the de-lamination occurs above \SI{0.58}{\percent}
and above \SI{0.62}{\percent} at \SI{4.2}{\kelvin}. On the whole
length, the samples are split between the buffer layer stack and the
superconducting layer indicating again a weakness of the interface
between these layers in the coated conductors from Fujikura. No de-lamination
is observed the other manufacturers' tapes during the stress - strain
measurements. In the stress - stain measurements, all samples are
as received without heat treatment, therefore any influence of the
mounting can be excluded.

\subsection{Normalized critical current vs. stress\label{sub:sub:Stess-dependence}}

\begin{figure*}[!tbph]
\begin{centering}
\includegraphics[width=0.5\textwidth]{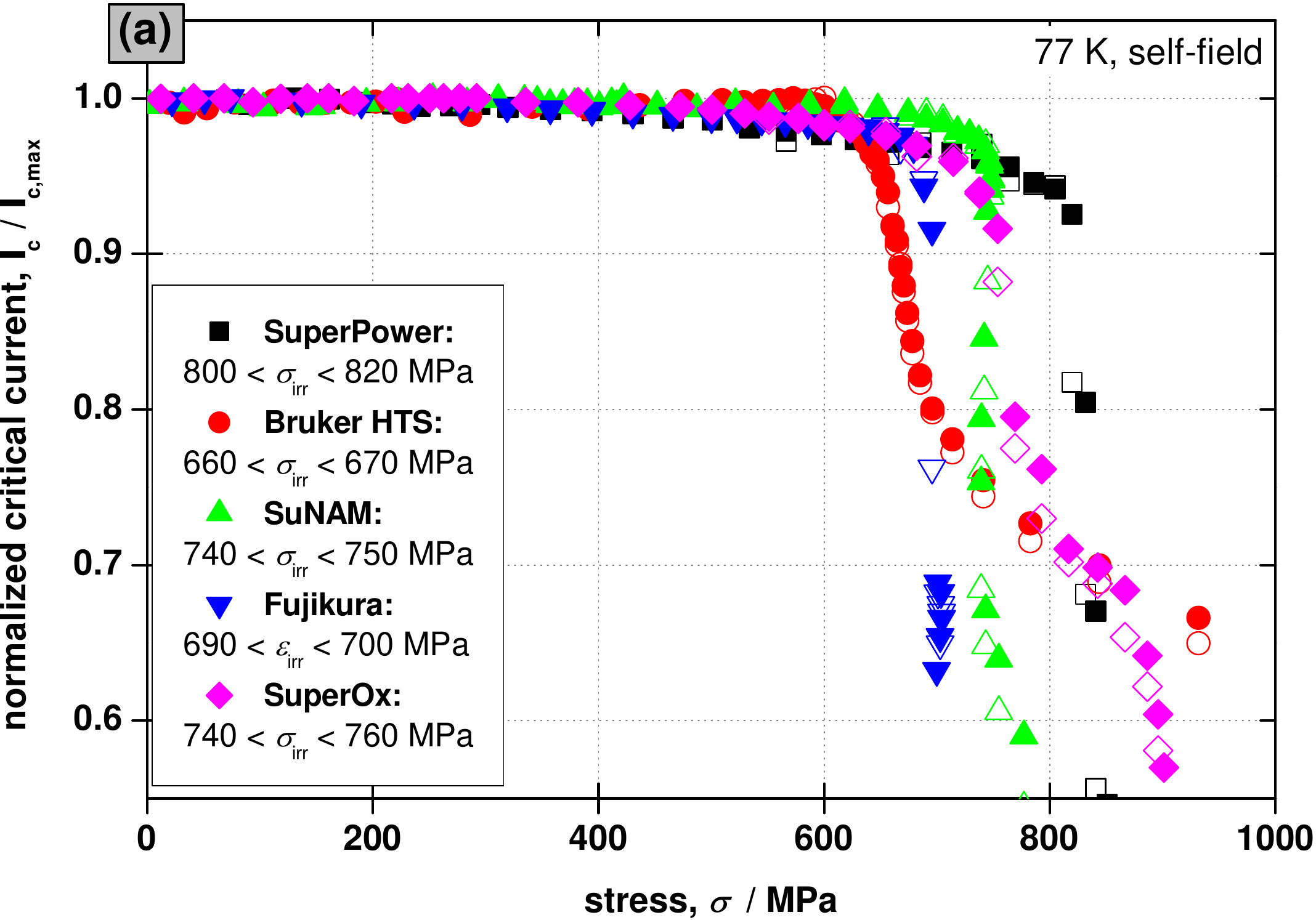}\includegraphics[width=0.5\textwidth]{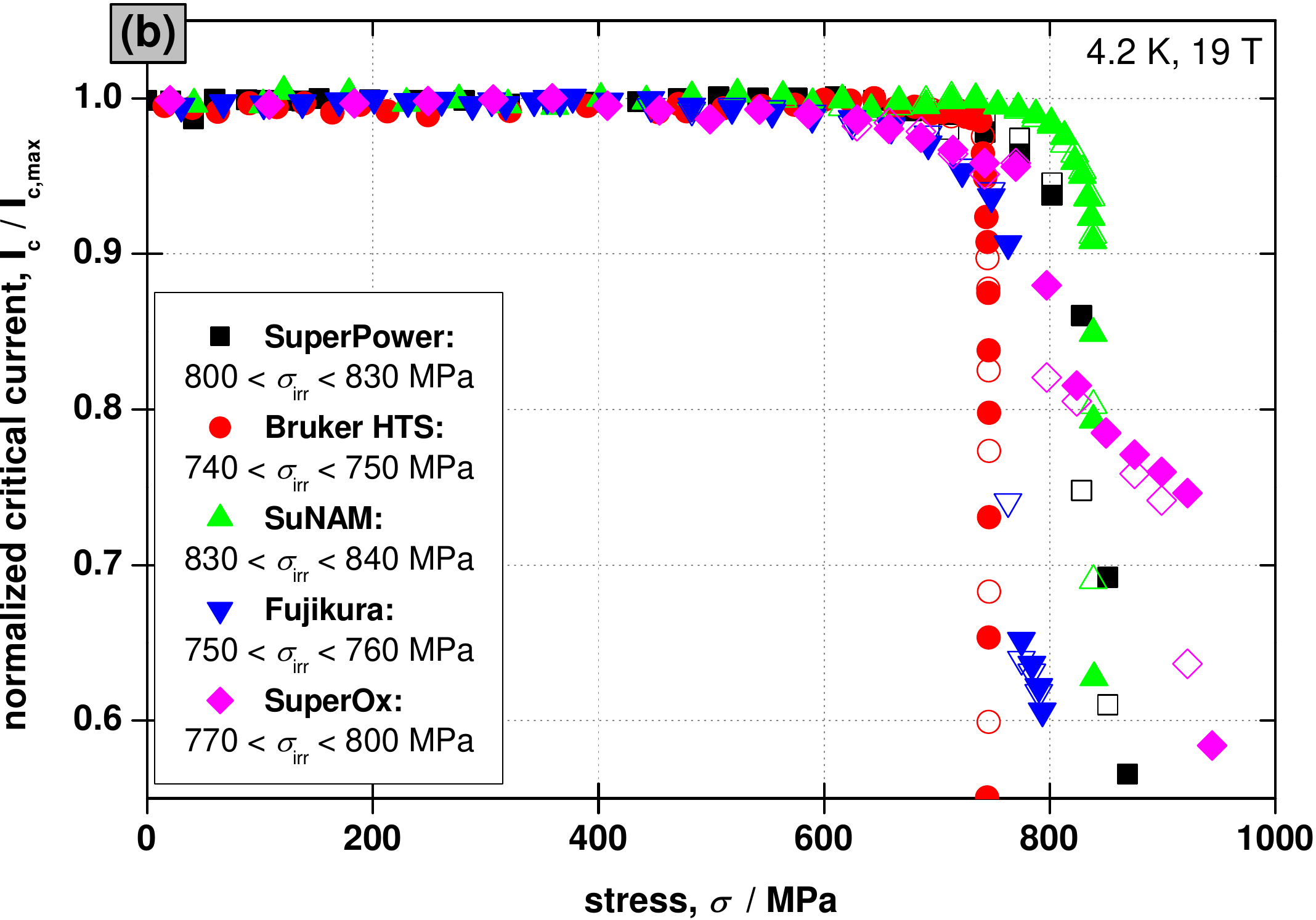}
\par\end{centering}

\caption{Stress dependence of various industrial REBCO tapes. Normalized critical
current $I_{\n{c}}/I_{\n{c,max}}$ versus stress $\sigma$ and irreversible
stress limits \sym{\sigma}{irr} at \SI{77}{\kelvin}, self-field
in graph (a) and at \SI{4.2}{\kelvin}, \SI{19}{\tesla} in graph
(b).\label{fig:Stress-dependence}}
\end{figure*}

Combining the strain dependent current carrying capabilities (subsection~\ref{sub:Strain-dependence})
with the samples' stress - strain correlation (subsection~\ref{sub:Stress-strain})
yields their normalized critical current versus mechanical stress.
The \SI{77}{\kelvin}, self-field stress dependence is shown in graph
(a) of figure~\ref{fig:Stress-dependence} and the \SI{4.2}{\kelvin},
\SI{19}{\tesla} stress dependence in graph (b). The differences between
the samples are lower in stress than in strain, as the tapes from
Bruker HTS which exhibit high irreversible strain \sym{\epsilon}{irr}
have low mechanical properties due to their stainless steel substrate,
leading to lower irreversible stress (\sym{\sigma}{irr} of \SIrange{660}{670}{\mega\pascal}
at \SI{77}{\kelvin} and \SIrange{740}{750}{\mega\pascal} at \SI{4.2}{\kelvin}).
The high mechanical strength of Hastelloy results in high irreversible
stress limits even for the tapes with low irreversible strain (\sym{\sigma}{irr}
of \SIrange{690}{700}{\mega\pascal}, \SIrange{740}{760}{\mega\pascal}
at \SI{77}{\kelvin} and \SIrange{690}{760}{\mega\pascal}, \SIrange{770}{800}{\mega\pascal}
at \SI{4.2}{\kelvin} for Fujikura respectively SuperOx tapes). This
brings the irreversible stress limits \sym{\sigma}{irr} of all investigated
REBCO tapes close together, especially at \SI{4.2}{\kelvin} where
all irreversible limits are within the \SIrange{740}{840}{\mega\pascal}
range.

\section{Discussion and conclusion\label{sec:Discussion-and-conclusion}}

The strain dependence of the tapes' normalized current carrying capabilities
shows strong differences between the manufacturers. The irreversible
strain limits range from \SI{0.45}{\percent} for SuperOx tapes to
\SI{0.72}{\percent} for Bruker HTS tapes. Due to similar conductor
layouts, the strain dependencies of tapes from SuperPower and SuNAM
are close, with irreversible strain limits of \SIrange{.66}{.69}{\percent}.
Fujikura and SuperOx tapes de-laminate during the electro-mechanical
measurements, resulting in a step like transition to irreversibility
and in lower irreversible strain limits. Neither de-lamination nor
such a step like transition is observed for any other manufacturer.
All samples were heated to temperatures \SI{10}{\percent} above the
manufacturers' recommendations during mounting, an impact of this
heat treatment cannot be excluded. Comparing \SI{77}{\kelvin}, self-field
and \SI{4.2}{\kelvin}, \SI{19}{\tesla}, the irreversible strain
limits of all samples are identical while the strain effect (the reversible
reduction of the critical current) is significantly lower at low temperature
and high field. Remaining below \SI{0.4}{\percent} strain, there
are no discernible differences in the samples\textquoteright{} strain
dependencies within the measurement accuracy. These results are in
good agreement with free-standing critical current versus strain measurements
which yield irreversible stress-limits \sym{\epsilon}{irr} of different
commercial conductors in the \SIrange{0.4}{0.8}{\percent} range \cite{Shin2007,Shin2012a,Shin2013}.
In a second step, the samples' mechanical properties are investigated
at \SI{77}{\kelvin} and at \SI{4.2}{\kelvin}. From the comparison
of the results at these two temperatures, we found for all the tapes
that the Young\textquoteright{}s modulus does not vary but the yield
strengths is ca. \SI{10}{\percent} higher at the lower temperature.
Young's moduli are in the \SIrange{149}{186}{\giga\pascal} range,
highest for Bruker HTS and lowest for SuperPower. All samples' curves
are nonlinear even at low strains due to the significant fraction
of copper which is always in yielding solely from the cool-down to
cryogenic temperatures \cite[p. 2-15]{EFDA-material-data-2007}. Bruker
HTS's curve is more round compared with the other samples because
of its stainless steel substrate making it the sample with the lowest
yield strength (ca. \SI{700}{\mega\pascal}). With ca. \SI{1000}{\mega\pascal},
the highest yield strength is observed for the REBCO tapes from SuperOx
mainly due to its strong Hastelloy substrate and low copper cross
section area. Full de-lamination is observed again with Fujikura REBCO
tapes during the mechanical measurements the sample are as received
without any heat treatment, thus indicating an interfacial (buffer
layer stack to REBCO layer) weakness. In a third step, the WASP measurements
are combined with the samples mechanical properties yielding the samples'
dependence of their reduced critical current on applied stress. While
the investigated coated conductor tapes are very different in their
irreversible strain limits, their irreversible stress behavior is
much more similar. Especially at \SI{4.2}{\kelvin}, \SI{19}{\tesla},
stresses have almost no effect on the current carrying capabilities
within the reversible region and the irreversible stress limits of
all samples are in the \SIrange{740}{840}{\mega\pascal} range. The
irreversible stress limits are lowest for Bruker HTS due its ``round''
stress - strain correlation combined with a low yield strength and
highest for SuperPower. Below \SI{600}{\mega\pascal}, there are no
differences in the samples' strain dependencies of their critical
currents. All mechanical and electro-mechanical properties of the
investigated REBCO tapes are summarized in table~\ref{tab:Summary}
at \SI{77}{\kelvin}, self-field and \SI{4.2}{\kelvin}, \SI{19}{\tesla}.

\begin{table*}[!tbph]
\caption{Summary of the electro-mechanical and mechanical properties of the
investigated REBCO coated conductors tape. \SI{77}{\kelvin}, self-field
data and \SI{4.2}{\kelvin}, \SI{19}{\tesla} data.\label{tab:Summary}}

\begin{centering}
{\tiny{}}%
\begin{tabular}{ccccc}
\toprule 
{\tiny{}{\tiny{}\SI{77}{\kelvin}}, self-field} & {\tiny{}Young's modulus, $E$} & {\tiny{}yield strength, \sym{R}{p0.2}} & {\tiny{}irreversible strain limit, \sym{\epsilon}{irr}} & {\tiny{}irreversible stress limit, \sym{\sigma}{irr}}\tabularnewline
\midrule
\midrule 
{\tiny{}Bruker HTS} & {\tiny{}\SI{186}{\giga\pascal}} & {\tiny{}\SI{696}{\mega\pascal}} & {\tiny{}\SIrange{0.70}{0.72}{\percent}} & {\tiny{}\SIrange{660}{670}{\mega\pascal}}\tabularnewline
\midrule 
{\tiny{}Fujikura} & {\tiny{}\SI{174}{\giga\pascal}} & {\tiny{}\SI{694}{\mega\pascal}} & {\tiny{}\SIrange{0.56}{0.58}{\percent}} & {\tiny{}\SIrange{690}{700}{\mega\pascal}}\tabularnewline
\midrule 
{\tiny{}SuNAM} & {\tiny{}\SI{168}{\giga\pascal}} & {\tiny{}\SI{755}{\mega\pascal}} & {\tiny{}\SIrange{0.66}{0.68}{\percent}} & {\tiny{}\SIrange{740}{750}{\mega\pascal}}\tabularnewline
\midrule 
{\tiny{}SuperOx} & {\tiny{}\SI{172}{\giga\pascal}} & {\tiny{}\SI{957}{\mega\pascal}} & {\tiny{}\SIrange{0.47}{0.49}{\percent}} & {\tiny{}\SIrange{740}{760}{\mega\pascal}}\tabularnewline
\midrule 
{\tiny{}SuperPower} & {\tiny{}\SI{149}{\giga\pascal}} & {\tiny{}\SI{868}{\mega\pascal}} & {\tiny{}\SIrange{0.66}{0.69}{\percent}} & {\tiny{}\SIrange{800}{820}{\mega\pascal}}\tabularnewline
\bottomrule
\end{tabular}
\par\end{centering}{\tiny \par}

\vspace{0.2cm}

\centering{}{\tiny{}}%
\begin{tabular}{ccccc}
\toprule 
{\tiny{}{\tiny{}\SI{4.2}{\kelvin}}, {\tiny{}\SI{19}{\tesla}}} & {\tiny{}Young's modulus, $E$} & {\tiny{}yield strength, \sym{R}{p0.2}} & {\tiny{}irreversible strain limit, \sym{\epsilon}{irr}} & {\tiny{}irreversible stress limit, \sym{\sigma}{irr}}\tabularnewline
\midrule
\midrule 
{\tiny{}Bruker HTS} & {\tiny{}\SI{187}{\giga\pascal}} & {\tiny{}\SI{736}{\mega\pascal}} & {\tiny{}\SIrange{0.70}{0.72}{\percent}} & {\tiny{}\SIrange{740}{750}{\mega\pascal}}\tabularnewline
\midrule 
{\tiny{}Fujikura} & {\tiny{}\SI{176}{\giga\pascal}} & {\tiny{}\SI{787}{\mega\pascal}} & {\tiny{}\SIrange{0.55}{0.57}{\percent}} & {\tiny{}\SIrange{750}{760}{\mega\pascal}}\tabularnewline
\midrule 
{\tiny{}SuNAM} & {\tiny{}\SI{168}{\giga\pascal}} & {\tiny{}\SI{846}{\mega\pascal}} & {\tiny{}\SIrange{0.67}{0.69}{\percent}} & {\tiny{}\SIrange{830}{840}{\mega\pascal}}\tabularnewline
\midrule 
{\tiny{}SuperOx} & {\tiny{}\SI{172}{\giga\pascal}} & {\tiny{}\SI{1083}{\mega\pascal}} & {\tiny{}\SIrange{0.45}{0.47}{\percent}} & {\tiny{}\SIrange{770}{800}{\mega\pascal}}\tabularnewline
\midrule 
{\tiny{}SuperPower} & {\tiny{}\SI{155}{\giga\pascal}} & {\tiny{}\SI{893}{\mega\pascal}} & {\tiny{}\SIrange{0.66}{0.68}{\percent}} & {\tiny{}\SIrange{800}{830}{\mega\pascal}}\tabularnewline
\bottomrule
\end{tabular}
\end{table*}

Below \SI{0.4}{\percent} strain and below \SI{600}{\mega\pascal}
longitudinal tensile stress, there are no differences between the
samples. On the other hand, in the case of applications where conductors
are exposed to higher strains or stresses, great attention needs to
be paid to the choice of the conductor. Common HTS applications are
within these limits (\SI{\approx220}{\mega\pascal} in Bi2223 NMR
coils \cite{Kiyoshi2011}, \SI{\approx330}{\mega\pascal} in HTS fusion magnets \cite{Bansal2008} and
\SI{\approx500}{\mega\pascal} in REBCO NMR coils \cite{Otsuka2010}) allowing
the conductor choice to be based on other properties as the transverse
stress dependence, the field dependence \cite{Senatore2014}, the
thermal conductivity \cite{Bonura2014a,Bonura2014b} or the electric
and the thermal stability. However, in small coils, combination of
hard and soft bending at the last turn of the layer can result in
strains above these limits and the strain sensitivity of the conductor
becomes the limiting factor.

\ack{}{}

The authors would like to acknowledge Marco Bonura for the coated
conductor micro-graphs, Davide Uglietti for his help with the the
zero strain determination, Andrea Lucarelli for his support with the
homogeneous soldering of REBCO tapes and Damien Zurmuehle for all
his help with the laboratory work and the experiments. Financial support
was provided by the Swiss National Science Foundation (Grant No. PP00P2\_144673
and Grant No. 51NF40-144613).

\bibliographystyle{unsrt}
\addcontentsline{toc}{section}{\refname}\bibliography{library}

\end{document}